# On-body Edge Computing through E-Textile Programmable Logic Array

**Frances Cleary[1,4*], David Henshall[2, 3], Sasitharan Balasubramaniam [1, 3]**

[1]Walton Institute, Waterford Institute of Technology, Waterford, Ireland

[2]Physiology & Medical Physics, Royal College of Surgeons Ireland, Dublin, Ireland

[3]FutureNeuro, The SFI Research Centre for Chronic and Rare Neurological Diseases, RCSI University of Medicine and Health Sciences, Dublin, Ireland

[4]Lero, the Irish Software Research Centre, Limerick, Ireland

\* **Correspondence:**
Corresponding Author
Frances.Cleary@waltoninstitute.ie

**E-textiles has received tremendous attention in recent years due to the capability of integrating sensors into a garment to provide high precision sensing of the human body. Besides sensing, a number of solutions for e-textile garments have also integrated wireless interfaces allowing these sensing data to be transmitted and also sensors that allow users to provide instructions through touching. While this has provided a new level of sensing that can result in unprecedented applications, there has been little attention placed on on-body computing for e-textiles. Facilitating computing on e-textiles can result in a new form of On-body Edge Computing, where sensor information are processed very close to the body before being transmitted to an external device or wireless access point. This form of computing can provide new security and data privacy capabilities and at the same time provide opportunities for new energy harvesting mechanisms to process the data through the garment. This paper proposes this concept through embroidered Programmable Logic Array (PLA) integrated into e-textiles. In the way that PLAs have programmable logic circuits by interconnecting different AND, NOT and OR gates, we propose e-textile based gates that are sewn into a garment and connected through conductive thread stitching. Two designs are proposed and this includes Single and Multi-Layered PLA. Experimental validations have been conducted at the individual gates as well as the entire PLA circuits to determine the voltage utilization as well as logic computing reliability. Our proposed approach can usher in a new form of On-Body Edge Computing for e-textile garments for future wearable technologies.**

## 1    Introduction

Electronic textiles (E-textiles) provides the ability to integrate electronic components and devices into garments [FER18] [CHI19]. Numerous solutions have been developed where sensors are integrated into the garment to do sensing that is close the body. Examples of these sensing capabilities includes electrocardiography (ECG) [RAM18], temperature [LUG18] and pressure [YAN19]. This technological development of integrating sensory devices into garments has resulted from advancement in nanotechnology materials, as well as micro-electronics, and traditional garment fabrication technology such as conductive threads and yarns [PAR16] [ISM20]. Numerous applications have resulted in sensor-based e-textile garments, including new techniques to monitor human physiological conditions as well as rehabilitation techniques and sports performance monitoring [KUB20].

**On-body Edge Computing through E-Textile Programmable Logic Array**

However, one capability that current e-textiles lack is embedded computing functionality to compute these sensor devices. Solutions have been proposed to have computing processors integrated into the garments through embroidering low-computational computing chips [ALH18]. These light chips usually have inbuilt Boolean logic circuitries that can process such sensor information, and in many cases also integrate wireless interfaces to enable transmission of the information to an external device [MEC15] [GON18]. However, there are a number of disadvantages with these current techniques around integrating these electronic-based chips into the garment. First, the form of the garment changes due to embroidering the electronic chips onto the fabric. The physical shape and protruding sharp edges on the chips can potentially lead to injury or discomfort of the user or other individuals in close proximity through scratching. Second, is the issue of security and privacy [ÖZL19]. Since these chips are programmable, they are prone to hacking where software changes on the chips lead to re-configuration of the logic circuits. Another issue to highlight in terms of privacy, is the fact that the raw data from the sensors are transmitted through wireless interfaces, and if made accessible to third party hackers this could result in detailed information of the user being exposed [BHU20]. This means that information pertaining to their physiological data such as temperature and pressure, as well as their health (for example, rehabilitation treatment progress) could be susceptible to being intercepted by an unauthorized and unknown entity [GOO18].

This paper proposes an alternative and new technique in conducting "On-body Edge Computing" [HAS19] [YU18]. We propose embroidering programmable capabilities directly into the garment to allow the Boolean logic circuits to be reconfigured to facilitate new applications or changes in the user's requirements providing a more personalized and safer approach to health monitoring at an individual level. The proposed approach uses the concept of Programmable Logic Array (PLA) found in typical electrically erasable programmable read-only memory (EEPROM) chips, and to embroider the logic gate circuit directly into garment. The programmable capability comes in the form of interchanging the stitching connections between the wired bus planes ( horizontal and vertical wires) from the different logic gates, in the very same way that the connections between the wired busses in the EEPROM can be re-programmed. Our proposed solution not only provides re-programmability by changing the stitching connections between the wired busses from the different gates, but also provides a layer of security by preventing hackers with accessibility to change the connections again through new stitching. Our proposed solution also creates a layer of privacy, where the computing is performed on the garment itself and through wired connections, where data from the sensors are computed through the logic circuits and the output can be a summary (e.g., "Yes" or "No") data that is transmitted wirelessly without leaving the detailed sensor data to be compromised by hackers.

Our paper presents a prototype of the "On-body computing" logic circuits through two varying design approaches 1) a single layer design 2) a multi-layered design. The paper does not go through the full solution of linking the sensor data to the logic circuits to be computed, but rather only focuses on the design and implementation of the embroidered logic gate circuitry and validation of the embroidered PLA performance. The performance analysis includes analyzing individual embroidered logic gate circuits as well as the overall embroidered PLA logic circuit. The overall logic circuit is analyzed through two different circuit model layouts to analyze the changes in the circuit analysis. Our proposed design through a multi-layered architecture leads to a reliable circuit that minimizes short-circuits between the wired bus planes and also enables easy re-configuration of the connections.

The paper is organized as follows: Section II presents the concept of PLA for e-textile garments. Section III discusses the development of individual logic gates, while section IV presents two designs of the PLA on the e-textile garment. Section V presents the results from the experimental analysis, and finally Section VI presents the conclusion.

## 2    PLA for E-textile Garment

Our proposed solution is illustrated in Figure. 1, where the PLA circuit is embroidered and stitched directly into the garment. We will first describe the concept of the PLA, the configuration of the PLA and enabling the Boolean Expression to align to two health focused use cases 1) Epilepsy and 2) Hypertension.

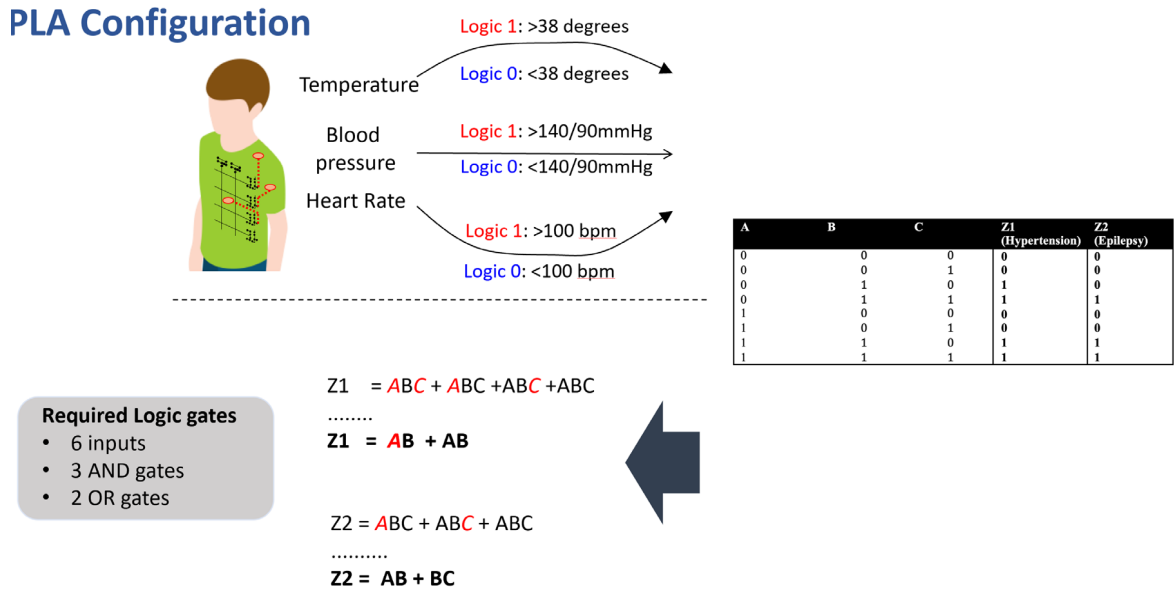

**Figure 1 PLA Configuration.**

PLAs are programmable logic devices that can create dynamic computing circuit logics. To enable the design and development of a reconfigurable "On-body edge computing" garment, the programmable logic array has a fixed architectural logic gate design along with programmable AND and OR logic gate capabilities through their respective embroidered connecting planes. The generic interconnected planes enables a combination of different logic circuits. The PLA is interconnected through multiple different planes, the AND gate plane, inverter and input planes and also the OR gate plane. Unlike a typical programmable array logic, the PLA enables all the gates to be programmed. This provides the required configurable digital circuitry architecture to meet the modular, personalised and reusable capabilities required for the "On-body Edge computing" garment. Taking this PLA architecture driven approach, we looked at the methodology and steps necessary to be completed in order to embed PLA technology within a garment structure utilizing embroidery techniques as an enabler.

### 2.1    Smart Garment PLA configuration

To map the embroidered PLA garment to a health sector use case, a truth table was defined to show the visual representation of the switching functionality of the PLA. A three-input truth table was adopted having 8 input combinations and two outputs represented by Z1 and Z2 logic levels. The truth table input combinations and output results were based on 2 health monitoring scenarios 1) Hypertension and 2) Epilepsy. Hypertension relates to high blood pressure levels in a human. Uncontrolled blood pressure can lead to serious health problems. Symptoms vary from irregular heartbeats to chest pain to difficulty breathing [FUC20]. Epilepsy is a common brain disease characterized by recurring seizures. Seizures are usually of two types, beginning in one part of the brain (localized, focal or partial seizures) or both brain hemispoheres simultaneously (generalized seizures). In the case of focal seizures, patients may or may not be aware of the start of the seizure. Symptoms can include sensory experiences (unusual taste, smell or epigastric sensation), as well as involuntary and repetitive movements of the trunk or arms. Seizures may also be associated with prodromal temperature fluctuations [DUB09] [SZE19]. The input variables in the smart garment PLA truth table are focused on representing sensor input readings linked to temperature, blood pressure and heart rate, where these variables align in certain cases to a raised alert notification. A defined threshold level for each sensor input reading was selected and set in order to represent logic level '1' and logic level '0' (Figure 1).

Taking the defined truth table, we deciphered the truth table Boolean Expression using the laws of Boolean algebra to reduce the logic function (Z1, Z2) to its *minimal form.* From the above computations (Figure 1) we can see that for the Garment PLA configuration we require 6 input connectors, 3 AND logic gates and 2 OR logic gates in order to design a PLA "On-body computing" garment.

## 2.2 On-Body Edge Computing PLA Challenges

The PLA that integrates multiple different gates, are digital logic circuits. However, in order to integrate them into e-textile garments, this will require converting the logic gates into embroidered circuits, that will produce digital outputs. The components will include electronic components (transistors and resistors), as well as conductive threads that will create the wired bus and the interconnection between the planes of the gates. A number of challenges will be faced in stitching the PLA into fabric. Firstly, energy will be dissipated at a very fast rate due to the circuitry to perform the logic gates. At the same time, this also depends on the arrangement of the circuits and their connections to minimize the path that will result in energy dissipation as the current is traveling between the gates. The next challenge lies in the efficient delivery of power as well as ground for each of the gates to ensure that this will not result in short circuits. The short circuits will result from overlapping that can occur if the wired bus starts to touch, and this can be based on certain movement and stretching of the garments.

## 3 Single Gate Model

Following the identification of the required number of core logic gates for the PLA configuration, the next steps involve the process of implementing the design of the required logic gates in an embroidered digitized format that can be sewn in using an embroidery machine into a garment's fabric back panel. The back panel of a garment was selected due to it being the largest area available for placement and manipulation of the logic gates. Here we detail the process utilised for the creation of individual gates and this includes the input connections (invertor logic gates), AND logic gates and OR logic gates.

In order to construct the logic gates the equivalent analog circuit was used as a base design guideline. The logic gates were designed based on RTL (Resistor-Transistor Switches), allowing the inputs to the logic gates to be connected directly to the base of the transistor switches. Through a combination of these embroidered RTL circuits with varying inputs, we can produce alert notification outputs linked to the hypertension and epilepsy health sensing use case truth table design described in section 2.

The graphic design of a logic gate embroidery circuit takes into consideration the space, position and placement of the resistors, transistors, power and ground connections and also the overall acceptable size of such a logic gate for it to be practically usable within a garment structure. The development of the embroidered digitized version of the circuit utilises an embroidery 'satin stitch' type for the creation of the embroidered stitching layout in the logic gate digitized design.

# On-body Edge Computing through E-Textile Programmable Logic Array

This above process was repeated in order to create the AND, OR and invertor logic gates prior to amalgamating them into the overall PLA circuit to be embedded into the garment. Table 1 provides an overview of the standard measurements used in the logic gate circuit embroidery design. Table 2 provides an overview of the components and materials used in the construction of the Single gate models and the design of the PLA's. The following sections provide an overview of the individual logic gates and their digitized embroidered circuit.

**Table 1  Logic Gate Measurements.**

| Design logic gate Properties | AND logic gate (single layer and multi-layer) | OR logic gate (single layer and multi-layer) | Invertor logic gate (single layer and multi-layer) |
|---|---|---|---|
| Embroidered Digitized gate width | 5.5cm | 6.5cm | 4.5cm |
| Embroidered Digitized gate length | 6.5cm | 6.5cm | 4.5cm |
| 1K Resistor connector nodes (length between) | 1cm | 1cm | 1cm |
| 10K Resistor connector nodes (length between) | 1.5cm | 1.5cm | 1.5cm |
| Transistor connector nodes (length between) | 2cm | 2cm | 2cm |

**Table 2 Component and Material Properties.**

| Component | Type | Modifications | Picture |
|---|---|---|---|
| Resistor (R2) | 1K resistor | Inclusion of end loops on their metal leads, in order to be sew onto the embroidered connection nodes points in the circuit | 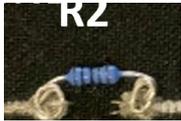 |
| Resistor (R) | 10K resistor | Inclusion of end loops on their metal leads, in order to be sew onto the embroidered connection nodes points in the circuit | 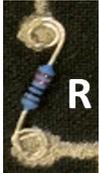 |
| Transistor | BC547B NPN Transistor | Inclusion of end loops on the connector, base and emitter transistor connections | 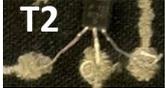 |
| Material Items | Type | Properties | |
| Conductive thread | Madeira HC12 Conductive with Resistance: <100Ω/m | Resistance: <100Ω/m | |
| Fabric Material | Smooth stretch Spandex (Black) | 82% nylon, 18% spandex | |
| Embroidery tear aware interfacing | Vlieseline Stitch n' Tear | Sewable, tear-off embroidery backing. 70% CV, 30% Cellulose | |

| | embroidery backing | 40 g/ m$^2$<br>Used as the main stabilizer for the embroidered logic gates. |
|---|---|---|

### 3.1    Embroidery digitized AND logic gate design.

The embroidery design of the AND logic gate had to accommodate 1x1K resistor, 2 x 10K resistors and 2 x BC547B NPN Transistors. Taking the standard RTL AND gate circuit as a base guide, an AND logic gate embroidery design is illustrated in Figure 2 (c).

The following specific design elements had to be considered and implemented from a functionality and efficiency point of view:

- Design of input terminals A, B connection points and selected shape allowing additional space on the fabric panel for interconnection of the AND plane horizontal and vertical lines. The extended shape and size of the input connectors also allow for ease of testing and measurements assessment, as the gate inputs can be clearly and quickly identified, and at the same time facilitate input measurements to be taken.
- Fixed placement of conductive connection nodes within the embroidered logic gates. Due to the need to attach resistors and transistors, it is important to include embroidered connector nodes for interconnection and placement of these components. Specific distances were selected and implemented to accommodate the components and these distances are shown in Table 1.
- From an overall circuit perspective, the clustering of more than one AND gate and their interconnections was another factor that had to be considered at the single gate design level. Each logic AND gate has its own GND and +Vcc connectors. Within the PLA we have a requirement for three AND gates to be interconnected to the AND plane, this meant that for every AND gate each had their own GND and +Vcc connection that had to be catered for. This was a challenging aspect to consider as it involved overlapping of various conductive circuitry thread elements that increased the risk of shorts occurring in the garment. In an attempt to limit the creation of such shorts within the overall circuit for the AND logic gate, we extended the GND connection embroidered line so that its connection aligned in parallel to the +Vcc connection to the left-hand side and this is illustrated in Fig. 3 (a) and (b). The GND connection is represented by an arrow and the +Vcc is represented by a line in Figure 2(d). The distance between the GND and +Vcc power lines is approximately 1cm to ensure no connectivity.
- Following individual validation of the operational functionality of the embroidered AND gates, a cluster group of 3 AND gates were combined (Figure 2 (e)), which was required for the overall circuit design.

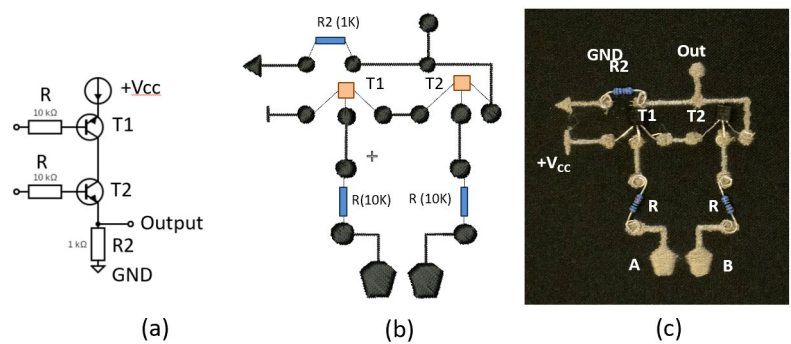

(a)                                          (b)                                          (c)

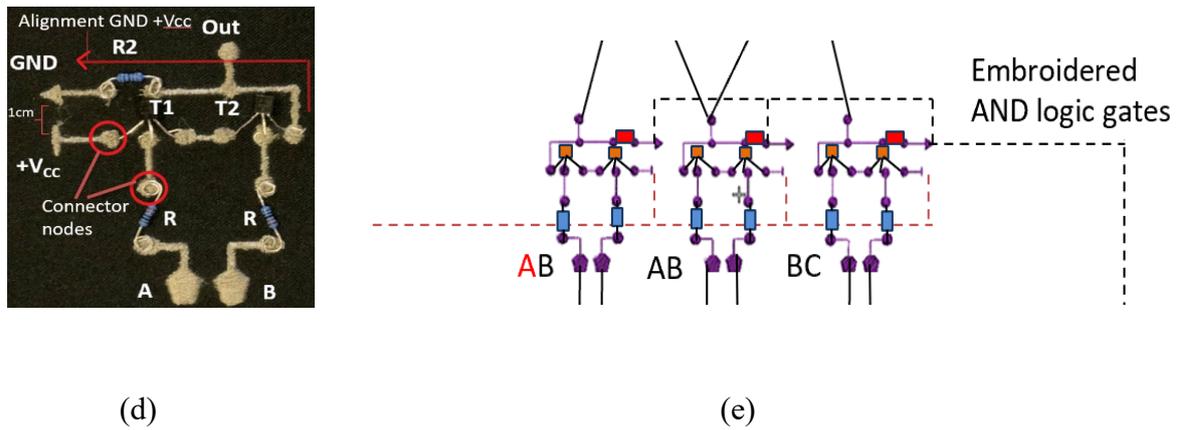

(d)                                                    (e)

**Figure 2  AND Logic Gate Embroidered digitization Design (a) 2-input logic AND gate can be constructed using RTL Resistor-transistor  (b) Embroidered digitized design for AND logic gate (c) Actual operational embroidered single AND logic gate  (d) AND gate embroidered design detailing connector nodes and modified GND conductive connection (e) structure of the AND gate cluster format.**

### 3.2    Embroidery digitized OR logic gate design.

Likewise for the OR logic gate, the embroidery design had to accommodate 1 x 1K resistor, 2 x 10K resistors and 2 x 2 x BC547B NPN Transistors.   Similar design input connections, conductive node connections and placement distances (Table 1) as per the AND log gate embroidery design were also adopted for the OR logic gate design to ensure the components fitted within the circuit's layout.

The following additional considerations for the design were considered for the OR gate:

- Due to the more complex and overlapping interconnections between the 2 transistors in the OR logic gate for the embroidered version, we instead looked at redirecting the transistor emitter embroidered connection under the 10K resistor and wrapped it around the outside of T2 in order to connect it to the output. This allowed us to reduce the number of overlapping wired bus and potential short circuits linked to this gate.
- By completing this redirection meant that this width of this gate was approximately 1cm larger than the AND gate (Table 1).
- By redirecting the connections also allowed for the GND and +Vcc to be strategically placed on the left-hand side of the OR gate, allowing for ease of connection when considering the 2 OR gates in the overall PLA circuit.
- More intricate level of connections were required to be embroidered into the OR gate, this meant that the OR logic gate width was slightly larger than the AND logic gate by ~1cm. The OR logic gate circuit design was modified allowing the power and ground connections to co-locate on the same side, therefore increasing efficiency and ease of connection.
- Only two OR logic gates were required due to the defined truth table containing two outputs (Z1,Z2). Hence his involved replication of two OR gates to be embroidered into the top half of the garment fabric at approximately 45 degree angles. 45 degrees angel was selected in order for ease of connection to the output node connection points coming from the cluster group of AND logic gates (Figure 3).

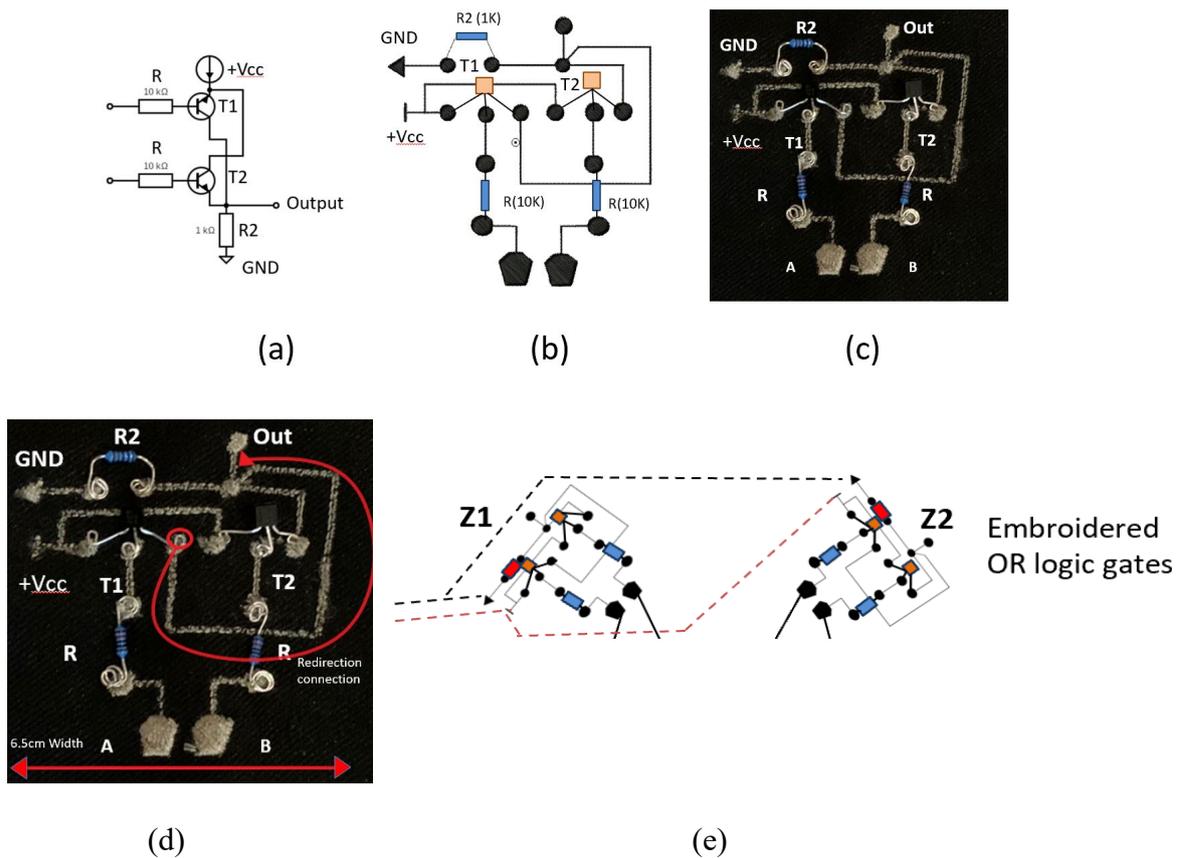

**Figure 3** OR gate embroidered digitization design (a) 2-input logic OR gate can be constructed using RTL Resistor-transistor (b) Embroidered digitized design for OR logic gate (c) Actual operational embroidered single OR logic gate(d) OR gate embroidered design modifications (e) OR gate cluster format.

### 3.3 Embroidery Digitized Invertor Logic Gate

The input connections to the PLA had the necessity to account for both the normal input and also the complemented input. In order to accommodate this need in the PLA from a design perspective we have to implement a NOT logic gate for every input. The NOT gate was constructed using RTL resistor transistor switching as a base. The function of this NOT logic gate was to provide the complement of the input signal.

The following design requirements for the Input/NOT logic gates had to be considered:

- The normal input and complemented input were kept as separate inputs for the sake of testing. We defined the circuit layout to have a standard input connector (e.g., representing 'A') and then a set representing 'complement A'. This design was taken for the other two input variables B and C, hence six input connectors were active as illustrated in Figure 4 (e).
- The length of the +Vcc and GND nodes were extended in order to allow for ease of connection to the other circuits GND and power lines again aligning them to the left-hand side of the design similar to the direction taken in the AND and OR gates above.
- Standard input connections (normal input) were placed vertically above the NOT embroidered parallel to the +Vcc connection. The +Vcc connector and the standard input connector were 4cm in length, and the GND connection was 4/5cm in length.

The GND connector length was a slight bit longer in order to allow for seamless connection between all GND connectors in the input panel.

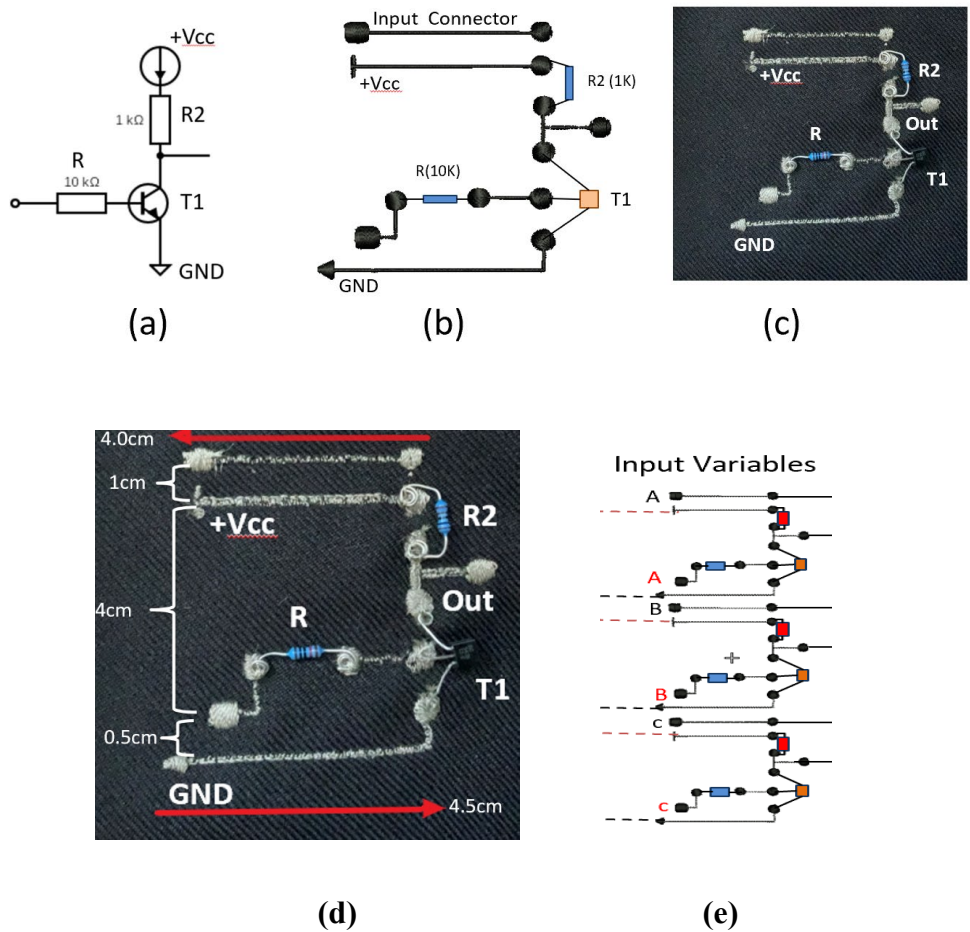

**Figure 4** Invertor Logic gate embroidered digitization design (a) 2-input logic OR gate can be constructed using RTL Resistor-transistor (b) Embroidered digitized design for OR logic gate (c) Actual operational embroidered single OR logic gate (d) Input/Or gate embroidered design modifications (e) Input/NOT gate cluster format.

## 4    Embroidered Digitized PLA Garment

Fusion of a PLA within an e-textile environment has not been investigated previously. Taking the modular individual embroidered gate and cluster gate groupings discussed in section 3, we now have the main building block in order to design the required PLA. To determine the best configuration of the PLA structure, we decided to develop two different designs, and this includes:

- Design 1: A *single layered approach* where all the Inputs/NOT, AND OR gates co-existed on one layer of fabric. This design fits to the traditional PLA structures seen from an electronics perspective.
- Design 2: A *multi-layered approach* consisting of 3 layers: 1) Input/NOT logic gates plane on the outer most layer, 2) the embroidered AND gate logic plane on the middle layer, and 3) the OR gate logic plane on the inside layer closet to the human body. All three layers will then be interconnected based on the defined Boolean expression and the physical overlay of the connecting programmable planes on the fabric layer.

It is important to note that when considering the fusion of the PLA into a garment structure the following elements had to be considered from the start:

- Practical placements of the embroidered logic gate circuit structures on the garment fabric based on the physical size of the required cluster of logic gates and location of inputs, outputs, GND and +Vcc locations.
- Based on the size of the back panel of fabric, it was necessary to limit the area and dimensions of the logic gates to a pragmatic size in order to be able to accommodate all the gates in a manageable format and practical layout on the fabric panel.
- PLA programmable requirements and alignment of the programmable planes based on the two different designs so the planes overlay (when required) and can interconnect allowing for stitched connections to be added. At the same time these interconnections should be easily removed or changed if required.
- Durability and appearance of the garment with inbuilt embroidered PLA 'on-the-body computing' functionality.
- Integration of the electronic components into the garment, using low resistance conductive thread, resistors and transistors.

## 4.1   Zig-Zag Wired Bus

Core to the workings of the PLA is the interconnection of the embroidered logic gate *planes* in order to be able to program and enable the e-textile PLA to have the expected computing functionality aligned to the working of the defined truth table use cases. For the two designs, different machine sewn planes were utilized in an attempt to analyze and compare their performance. For the single layer design, 'zig-zag' stitching in a horizontal line was used and this is illustrated in Figure 5. The zig zag stitch type was selected as it is the most suited stitch type to use in conjunction with a lycra based 4 way stretch material. Each horizontal plane zig-zag plane wired bus linked to an input reached approximately 14.5cm in length.

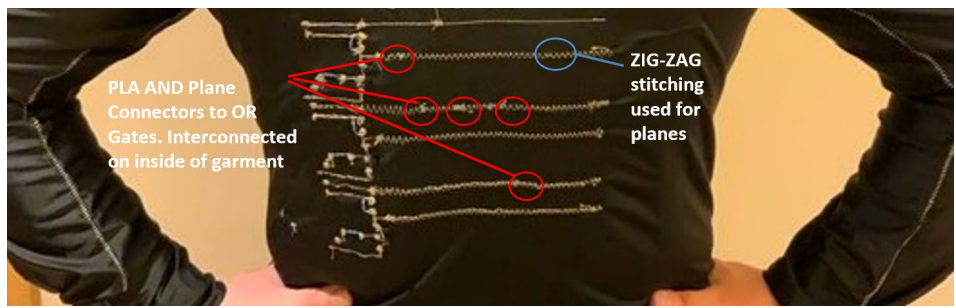

**Figure 5  Zig-Zag wired bus stitching plane for design 1.**

## 4.2   Design 1: Single Layered PLA

The focus of the Single layered PLA design was to check the viability to see if the e-textile PLA could be incorporated on a single back panel layer of fabric. Layout of the logic gate devices based on the criteria defined in section 3 was a vital activity allowing for creative experimentation and determining best placement on the fabric. White tailor chalk was used to mark the placement of the logic gate positions, prior to hooping the fabric with tear away interface backing in the embroidery hoop, this was required as part of the setup steps before using the embroidery machine. The full circuit layout for design 1 can be viewed in Figure 6 (a).

From a design perspective the cluster group of Input/NOT variable connectors were placed in a vertical manner towards the lower left side of the garment. The AND logic gate clustering grouping was placed in the middle of the garment allowing for the AND plane vertical connections (AB, AB, BC) to be interconnected to the Input/NOT plane horizontal connections shown in Figure 6 (b).

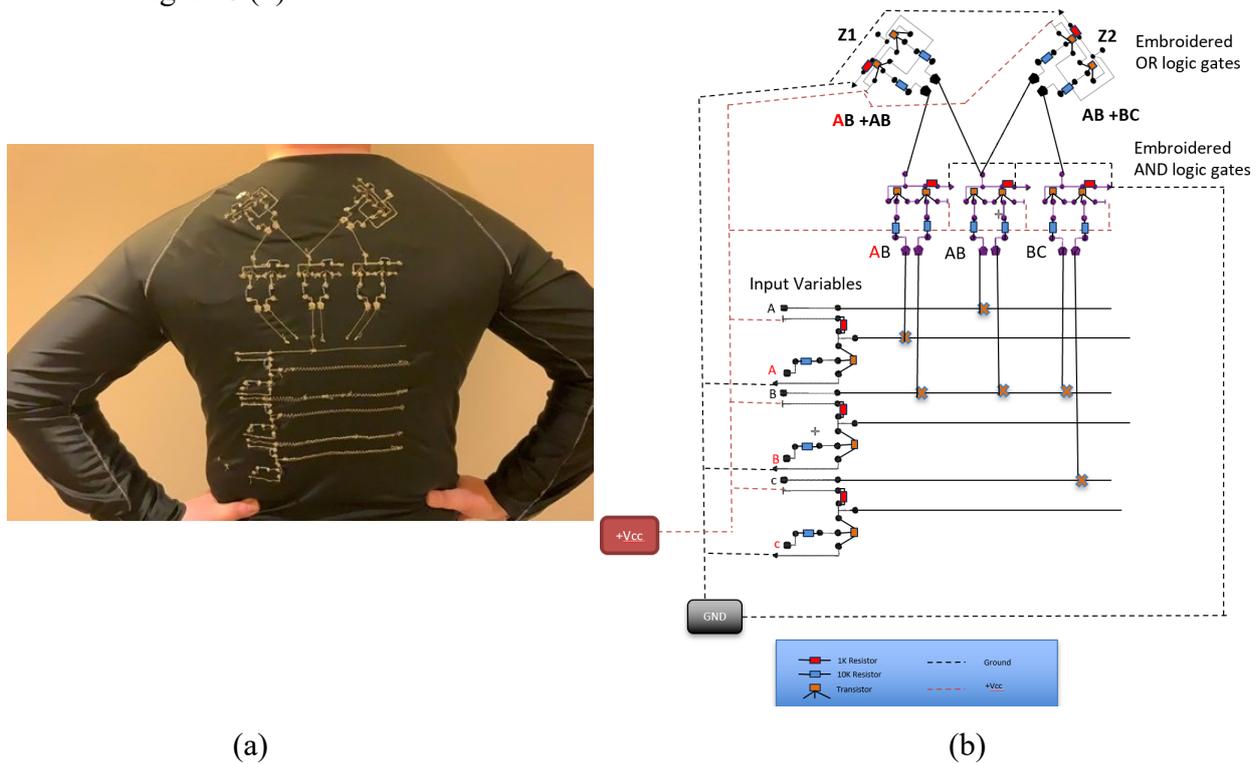

(a)                                                                                              (b)

**Figure 6  Single Layer E-textile 'On the body worn computing' PLA garment (a) E-Textile Single Layer PLA garment (b) Single Layer embroidered circuitry overview.**

Outputs from the AND logic gates were connected directly to the embroidered OR logic gate inputs allowing the final computational switching functions to occur. In order to be able to determine if an output was a high or low level alert notification, an LED was connected to both OR gate outputs Z1 and Z2. The OR gate output result could then be cross compared with the initial hypertension or epilepsy health monitoring truth table to determine the meaning of the result.

The following challenges were encountered in the initial single layer PLA garment. Learning from these identified challenges we then proceeded to progress towards design 2 to develop a more efficient design.

1.  Due to each logic gate having both a GND and +Vcc connection, it was difficult to create a logical GND and +Vcc circuit flow that could cater for all the necessary interconnections in this single layered design. The high volume of connections between the gates and lack of insulation also lead to many short circuits occurring, and the performance of the garment was reduced.

2.  In this version, the AND logic gates logic GND and +Vcc were pointing to the right in the architecture design. However, in the Multi-layer design 2, we reversed this so the GND and +Vcc were going in the left direction to enable the interconnection of the GND's and +Vcc's together for greater efficiency.

3.  The required interconnections from the AND gate Plane to the OR gates inputs was created using individual pieces of conductive thread on fabric strips at the back of the garment. This was not an ideal solution as the fabric connections were prone to a lot of movement and created many shorts and hence greatly reduced the efficiency of the garment.

4.    On a number of occasions a negative flow of voltage was detected between the OR and gate connections. This was most likely as a result of the positive side of the circuit and the negative terminal side of the circuit being connected at a point in time, altering the voltage orientation in a circuit.

## 4.3    Design 2: Multi -Layered PLA

Design 2 took an alternative approach of layering the modular logic gate elements of the PLA into the garment, and this is illustrated in Figure 7, where the PLA has three planes each containing specific types of gates that needed to be interconnected. The three layers include:

- Layer 1: Input/NOT logic gates plane on the outer most external layer of the garment.
- Layer 2: AND gate logic plane is housed on the middle fabric layer.
- Layer 3: The OR gate logic plane is housed in the inside fabric layer closet to the human skin.

The purpose of this 3 layered design approach is to overlay the layers in a manner where we can allow for the programmable interconnections from one plane to another plane, and this is achieved by stitching the interconnections. Also the length of the vertical/horizontal lines was decided based on the plane overlays and the most appropriate placement position on the garment in order to improve the embroidered PLA's efficiency and voltage performance. Hence the angle/width/direction of the plane in each layer incorporates a mixture of vertical and horizontal lines as shown in Figure 7 (a) – (c). The planes were sewn using a 'straight running machine stitch' onto specific marked sections of the three fabric layers so that they could interconnect with ease and allow for the necessary programmable stitching to be completed.

Initially, layer 1 was implemented, where the Inputs/NOT logic gates were embroidered onto the top layer of fabric. The Input plane connections for this design was placed in a vertical manner down the left-hand side of the fabric as shown in Figure 7 (a). The line lengths varied from 15cm to 24cm in length depending on the input position in the input/NOT gate logic cluster group.

Layer 2 involved planning the placement of the AND logic cluster group so that the gate inputs could interconnect and fuse to the input vertical plane connections coming from layer 1.  On the right-hand side of the AND logic gates where the outputs reside, approximately 9cm space was allocated to include the gate's output connection plane that was required to link to the OR logic gate plane (layer 3), as shown in Figure 7 (b). On the left-hand side of the AND logic gate, approximately 12cm space was required to be allocated in order to align the AND plane with the input plane on layer 2. A combination of vertical and horizontal lines were used to efficiently line up the planes enabling easy stitched connection of the required variables across the 2 layers.

Layer 3 consists of the two OR logic gates representing the outputs Z1 and Z2, as shown in Figure 7 (c). Layer 3 is the closest to the human skin, so we selected the cluster group with the least number of gates to reside on this layer due to the close contact with the skin. It is worth noting that for this layered approach, apart from the tear away interface stabilizer that was used to support the embroidered gates, an iron on fusible backing was also applied to the under layered part of the garment. This was beneficial as it reduced the possibility of any short circuits occurring (e.g., conductive threads touching) and it was also soft to the touch, hence preventing any irritation against the skin.

Once all the individual layer planes were embordered into the garment, the programmable connections were then sewn in to link all the layers as illustrated in Figure 7(d). Figure 7 (e) shows the design with all the different layers. Interconnections were completed using conductive thread to stitch the corresponding plane variables together as per the defined Boolean expression computed in section 3. Figure 7 (f) shows the connections that were stitched into the embroidered PLA in order to link the modular circuit layers together.

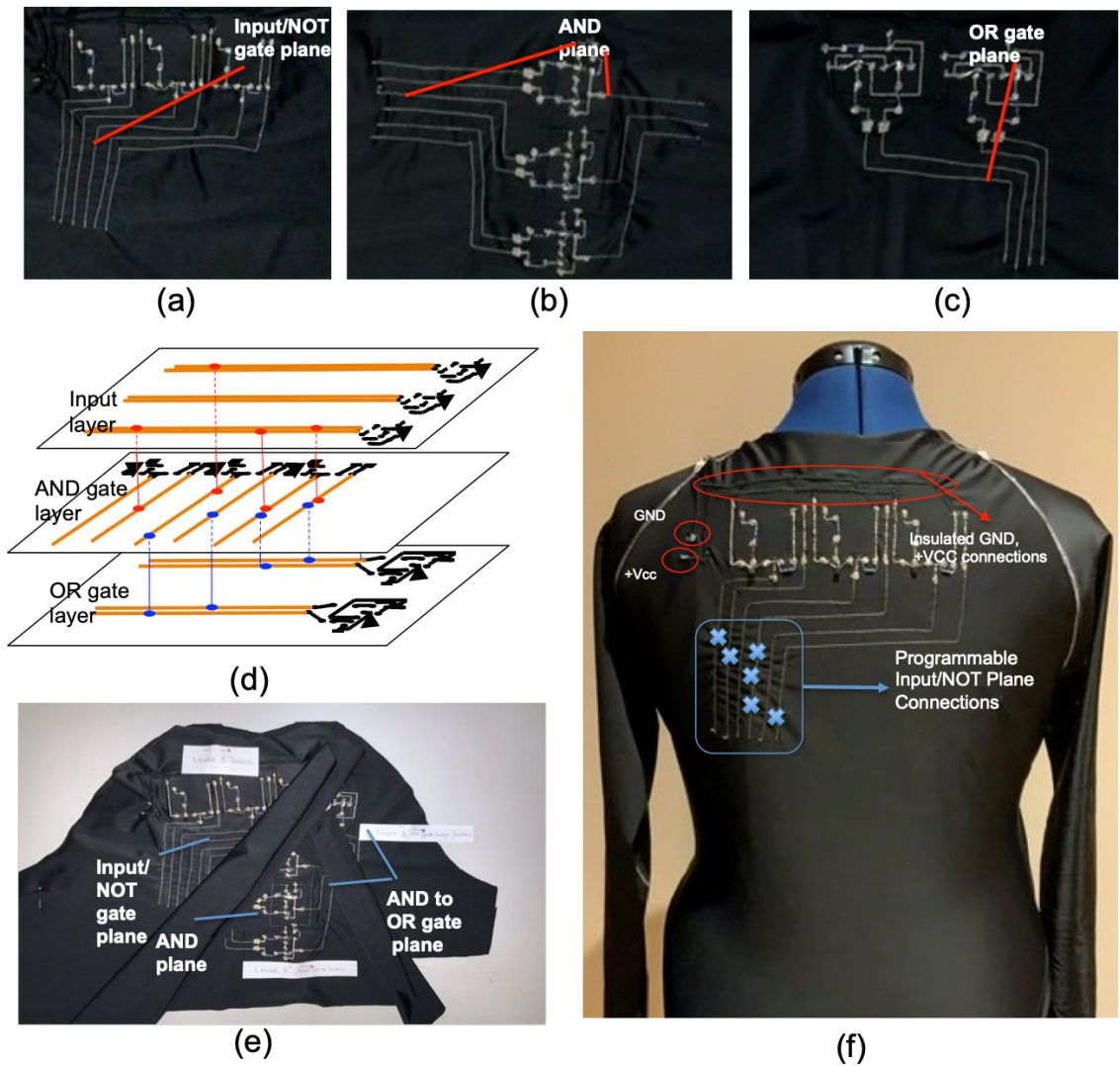

**Figure 7  Multilayered PLA Smart Garment (a) Input/NOT embroidered logic gate cluster and plane (b) AND embroidered logic gate cluster and plane (c) OR gate logic gate cluster and plane.**

With regards to +Vcc and GND connections required for each of the logic gates, this was where we learnt from the single layered design and adapted our design approach for the multi-layered design. All power and GND connections were directed towards the left-hand side of the garment as shown in Figure 7 (f). This was completed on all three layers, then all GND's and +Vcc were insulated by using a compressed zig-zag machine stitch as shown in Figure 8 (c) encapsulating the conductive thread path and preventing it from coming in contact with any other conductive elements to potentially create a short circuit in the embroidered PLA.

With all the power and GND connections located at the left-hand side of the garment, a layered connection was completed, linking all the 3 layers and creating a more solid and reliable one point location for the battery connection to the top of layer 1 to connect to provide the required ground and power source.

## 5    Results and Performance Analysis

The development of the two smart garment PLA designs, provided us with the opportunity to assess the performance of each individually and also cross compare from an efficiency and performance point of view.

This section will analyze the measurement performance of the circuit by focusing on the (i) Input/Not connector plane, (ii) embroidered AND logic gate, (ii) embroidered OR logic gate, (iii) PLA plane interconnection, and (iv) comparison between the Single l and Multilayer designs for the full e-textile PLA on-body edge computing.

A systematic approach was taken to initially check the workings of each logic gate on the garment of both designs before validating the PLA e-textile embroidered circuit as a whole. The key measurement that was focused on was the voltage reading measurements across the circuit designs. For each design assessment, the input/NOT, AND and OR embroidered logic gates were validated individually to observe their functional operations (input and output voltage readings) when power was applied to the circuit. In order to be able to determine a high level '1' output coming from the OR logic gates, LEDs were included into the design on each PLA (Figure 8 (a-b). This enabled us to actively step through and validate the defined truth table variable inputs, apply the corresponding input to the PLA input connectors and assess the outputs from the embroidered logic gates (Z1, Z2). This section will highlight our experimental validation and analyze their outcomes.

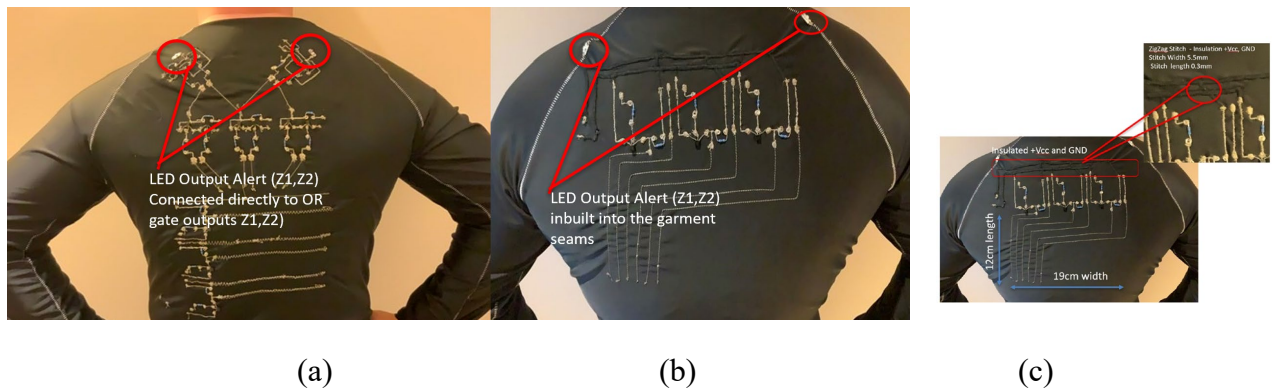

(a)                                   (b)                                   (c)

**Figure 8  LED alert notification included into the garment raising alerts when an output high level is computed (a) Single layer LED Z1, Z2 alerts (b) Multi-layer LED Z1, Z2 alerts (c) +VCC and GND insulation compressed zig-zag stitch and Input/Not Plane dimensions.**

## 5.1   Input/NOT Connector Plane Validation

Due to the varying design elements incorporated into the single layered design and the multi-layered design, the input/NOT plane had varying stitching types and layout options as detailed in section 4.1 and 4.3. In order to validate the computing operation of the inputs, we stepped through the truth table inputs and observed the voltage readings present at the end-most point of the input/NOT planes for both designs.

Using a standard 9-volt battery to power the garment, each individual input connection was validated to ensure the input voltage readings met the desired measurements. As can be seen from the sample input connector graphs in Figure 9 (a-b), there is little variation in the voltage obtained from the input/NOT connector plane in single layer compared to the multi-layer designs. On average the input levels for the normal input connection ranged from 8.02 volts to 8.67 volts, when the input connections to A, B, C were in a high-level state of '1'. When the complement input connections A, B, C for the NOT logic gates were supplied with power, the end plane voltage reading ranged from 0 to 0.43 Volts. From the tests completed, it can be concluded that due to the varying length of the planes and the different types of stitching types used in the designs, the end result had little impact on the overall efficiency of the embroidered circuitry functional operational aspects of the garments. Purely from a practical garment design and performance point of view, seeing as there is little difference in the voltage different, the option to use a zig-zag would be a preferred design method as it allows for additional levels of moveability and flexibility in the garment from a usability and wearability point of view. The Zig-zag stitching when stretched could expand from 14.5cm to 19.5cm on the garment. Having this expandability enables an additional level of in-built flexibility in the garment in order to be able to cater for varying body sizes of end users.

## 5.2    Embroidered AND Logic Gate Validation

The cluster group of AND gates were placed in different locations depending on single layer or multi-layer design and each had different methods of overlaying and connecting to the required programmable planes as detailed in section 4.2 and 4.3. The AND logic gates are central to the overall PLA, and the analysis of the voltage out of the embroidered AND logic gates is vitally important as this gives us an insight into the operational functionality of the gate. This is to determine if the input voltage is high enough to trigger the necessary output voltage level threshold to feed into the other embroidered OR logic gate ensuring the PLA operates correctly.

Assessment of voltage readings taken from the AND Gate plane were analyzed based on the voltage readings at the 2 terminal inputs for each AND gate. In total in the PLA, we have 3 individual AND gates connected to the Input connection Plane described in the previous sub-section. Figure 10 (a) (b) provides an example of the voltage readings taken at AND gate 2 for each input based on the applied truth table input variable readings (i.e., 000 to 111).

For both designs only slight variations in voltage readings were detected at the AND gate input connections, this was mostly due to the variation in the actual input source battery supply (i.e., for a 9-volt battery after usage for various iterations of testing, its actual voltage could dip to between 8 to 9 Volts) this in turn lead to slight variations of inputs at the AND gates. This concludes that the connections from the Input/Not connector plane to the AND connector planes on both designs worked as expected.

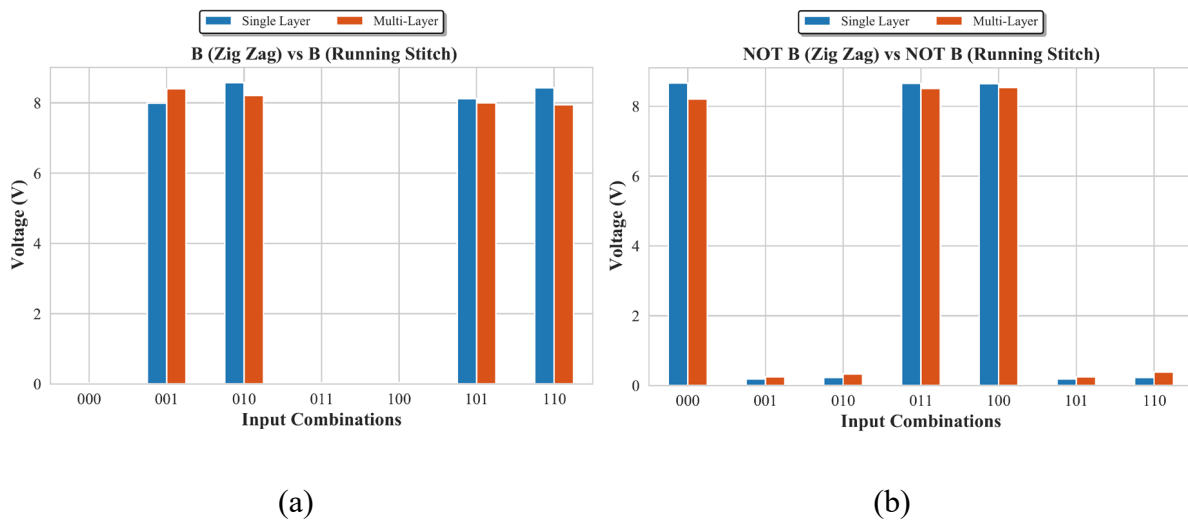

(a)                                                    (b)

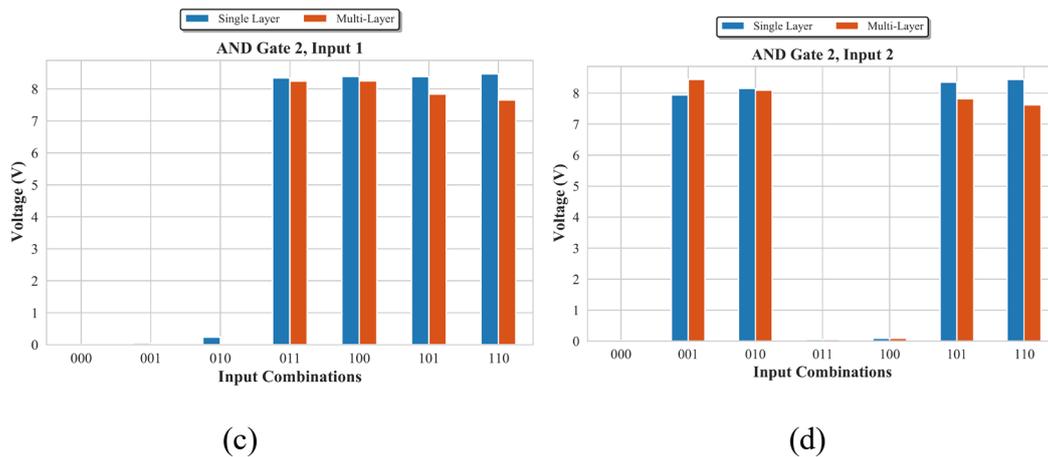

(c)                                          (d)

**Figure 9  Graphs representing the voltage output readings coming from the Input/NOT logic gate connectors and Embroidered AND logic gate number 2 input voltage readings (a) Comparison voltage readings B Input connector (b) Comparison voltage readings NOT B Input connector. (c) AND Gate 2, Input 1 voltage readings (d) AND Gate 2, Input 2 voltage readings.**

The voltage output readings from the AND gates demonstrate a substantial difference in the values as shown in Figure 10 (a) Gate 1, (b) Gate 2 and (c) Gate 3. We observed that there was a two-fold increase in the voltage levels exiting the AND gates in the Multi-layered PLA design in comparison to the voltage readings coming from the Single layer PLA design. Based on readings obtained on average the AND gate output for the single layer gate had a reading of 2.7 Volt whereas the Multi-layer AND gate output had an average high-level reading of 7.04 Volts.  The reason for this could be linked to a number of factors that are associated to the embroidered PLA circuitry.

- In the case of the single layer design, the AND gates are embroidered in a vertical pattern across the middle to high part of the garment fabric, whereas the AND gates in the Multi-layered design are embroidered in a horizontal manner across the middle of the fabric back panel. The layout utilized in the Multi-layered approach enables a stronger voltage reading to be emitted from the output of the three AND gates. Based on the voltage readings, we can clearly decipher that there was enhanced current flow in the Multi-layer AND gate cluster that was embroidered in a horizontal layout position on the fabric as opposed to the single layered embroidered AND gate cluster that was in a vertical position on the top middle level of the fabric back panel.
- On the single layer, the AND gate cluster design and layout had the GND and +VCC connections facing to the right side of the garment, This meant that there was connection overlays of GND and +Vcc required in order to interconnect all together. To accommodate this the GND and +Vcc connecting interfaces were sewn in using conductive thread on strips of fabric in order to create the required bridge connections. This approach led to a lot of movement of the fabric strips on the inside of the garment and hence introduced random short circuits. In addition to this the GND and +Vcc connections were not insulated and this also led to short circuiting more in the Single layer garment as opposed to the multi-layer garment. This is turn caused a lot of operational challenges for the cluster of AND gates within this design.

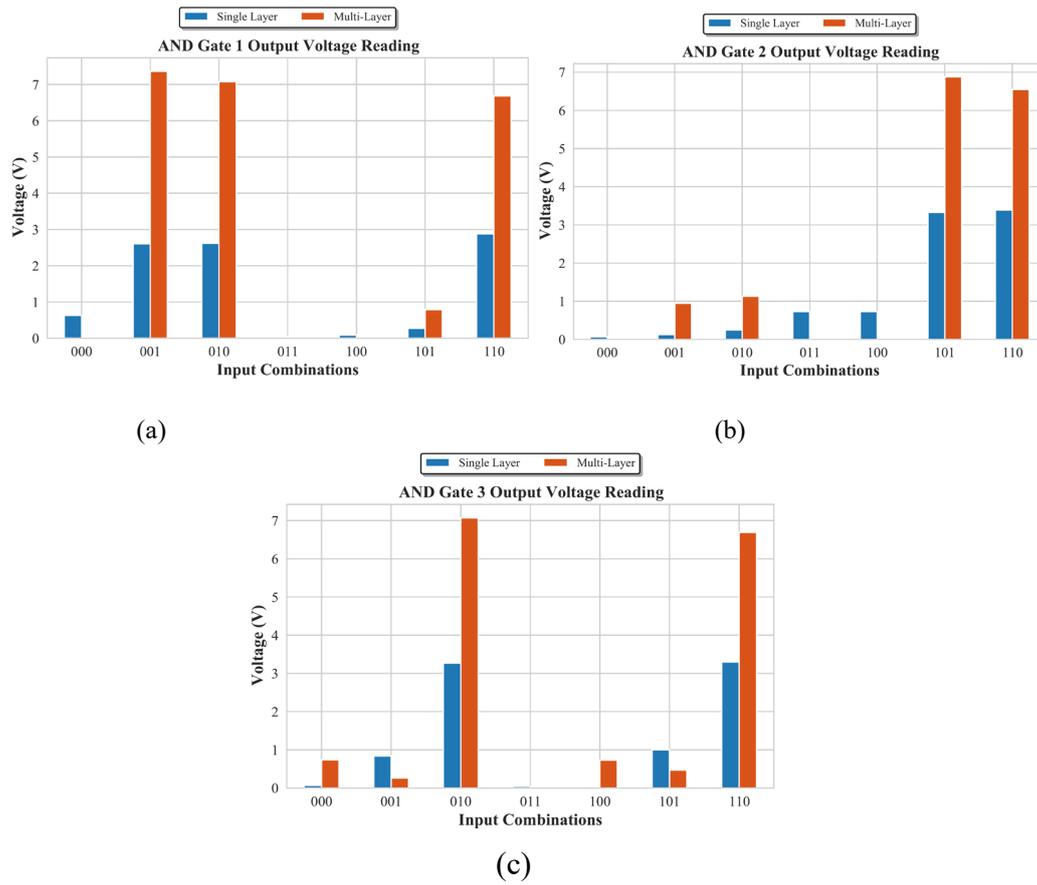

(a)                                            (b)

(c)

**Figure 10   Comparison output voltage readings comparing the design of Single layer and Multi-layer AND logic gates for (a) Gate 1, (b) Gate 2, and (c) Gate 3.**

### 5.3   Embroidered OR Logic Gate Validation

The voltage readings coming from the AND gates plane discussed in section 5.2 has a direct impact on the embroidered OR gate input and output voltage readings. For the single layered PLA garment based on the decreased voltage readings coming from the AND logic gate outputs as discussed in section 5.1.2 when this in turn was fed into the programmed OR gates based on the combinational truth table logic and Boolean expression being applied (Figure 1) resulted in a noticeable decrease in the output voltage levels from the embroidered OR logic gates as shown in Figure 11. In comparison the Multi-layer OR gate output readings (Z1, Z2) were on average double in the voltage level. The voltage depletion across the AND and OR gates in the Single layer design raised challenges linked to powering additional alert notification elements connected to the OR gate output (e.g., LED). For a high-level output from the Single layer design garment, the voltage reading fluctuated between 1.91 to 2.58 volts whereas in comparison for the Multi-layered design it fluctuated between 4.31 to 5.1 volts.

During testing a noticeable LED light reduction was experienced while testing the truth table configurations to ensure the correct outputs were being computed and achieved. The multi-layered design approach with interconnected programmable layers and also insulated +Vcc and Gnd connections proved to be a more efficient and higher performing smart PLA garment as opposed to the single layered design PLA garment.

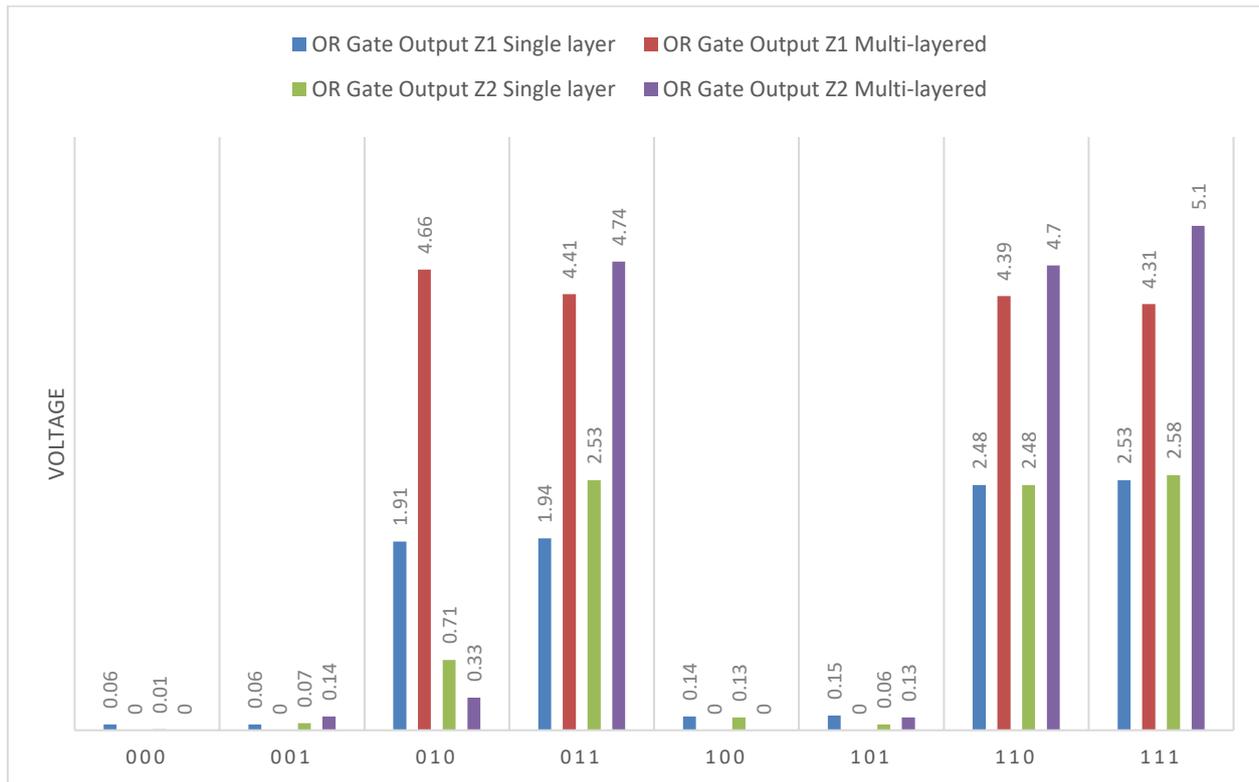

**Figure 11    Output OR gate Z1, Z2 Single-layer versus Multi-layer PLA garment.**

**5.4    Single layer and Multi-layer validation of the overall e-textile PLA garment**

Once all the elements were embroidered onto the fabric, the components added and cross checked, then the connections were stitched in for the programmable connections across the required planes, and this was the programmable aspect of our embroidered PLA. Once this was completed standard continuity testing was completed to:

1.    Ensure the Input/NOT connections (A, B, C) connected successfully to the inputs of the relevant AND gate connections as detailed in Figure 12 (c).
2.    Identify and rectify any short circuits detected across the planes.

Following this initial continuity test and validation, the PLA embroidered fabric panels were sewn into the full garment version and overall testing of the system was undertaken based on the defined truth table in Figure 1. This testing is based on applying the inputs to both of the garment designs input connectors and visually confirming if the correct alert notifications were raised, as per the defined outputs Z1, Z2 on the truth table. This was completed multiple times across both garments, and Figure 12 (a -b) shows the LED output that will indicate a correct logic circuit output (red LED is for Z1 output, while blue LED is for Z2 output).

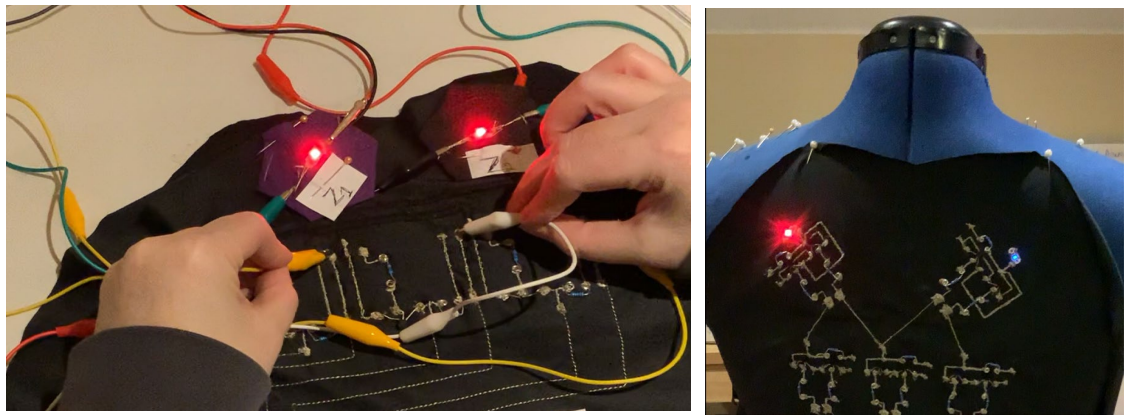

(a)          (b)

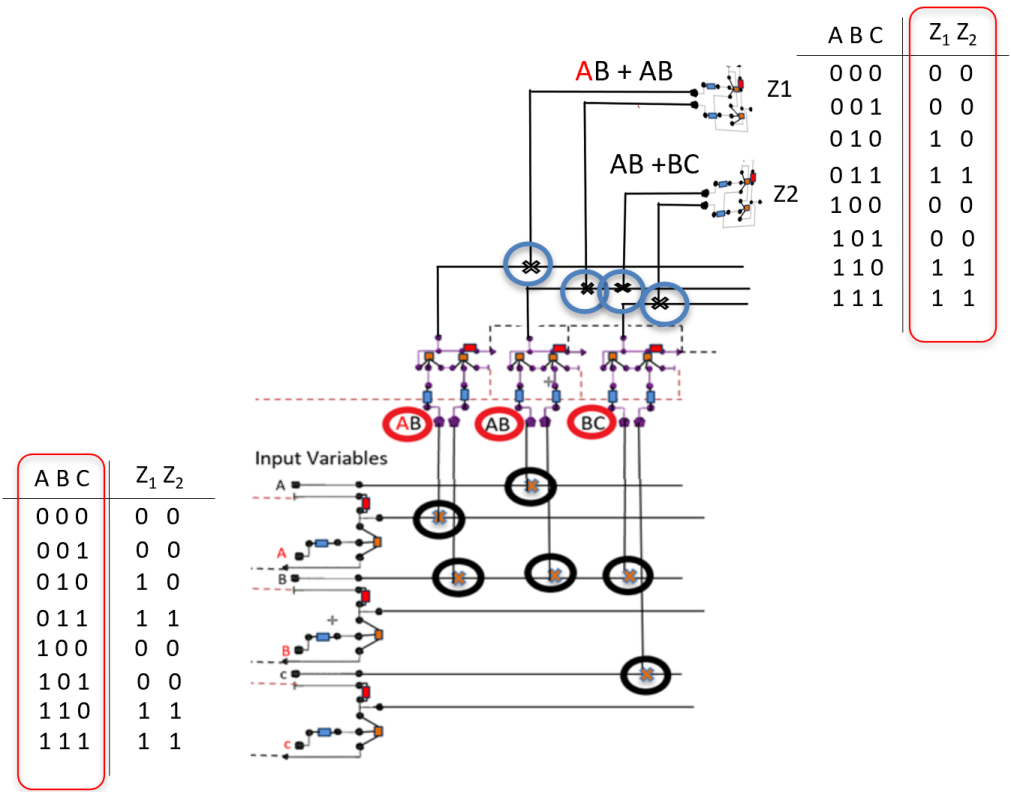

(c)

**Figure 12  Validation of the 2 PLA smart Garment Designs with LED outputs indicating a positive result of the logic circuit through OR gate outputs Z1,Z2 (a) Single -Layer  (b) Multi-Layer (c) Programmable PLA smart garment connections.**

Table 3 presents the true and false output comparisons between the single and multi-layered designs. Based in the results coming from these tables, True outputs mapped to True positive (the outputs achieved were the correct results), Indeterminate mapped to True Negative ( the outputs obtained was the wrong results and either no led or the wrong led was lit up), False Outputs mapped to False Positives ( this was where a short circuit was experienced and incorrect false output results was observed).

**Table 3: Comparison of Single-Layer and Multi-layer true and false output results.**

| Output Result | True Positive | True Negative | False Positive |
|---|---|---|---|
| **Single-Layer (No Stretch)** | 26 | 11 | 3 |
| **Single-Layer (Stretch)** | 21 | 19 | 0 |
| **Multi-layer( No Stretch)** | 37 | 0 | 3 |
| **Multi-layer (Stretch)** | 32 | 0 | 0 |

Based on the results in Table 3, when testing multiple iterations of the single layer in the garment, a large number of indeterminate output results were obtained, the reason for this is that the voltage level that was reaching the OR logic gates was not high enough to generate sufficient current to power the LED linked to the output of the OR gate. For a number of the cases where the Z1 and/or Z2 output were at level high '1', we can see that the output voltage fluctuated between 1.7 Volts to 2.06 Volts. Within the garment this initially powered the output from the Z1 OR gate (red LED) but over time the light intensity of the LED diminished to a level where it was difficult to read. The Z2 (blue LED) struggled to maintain any strong visible light alert throughout the tests. This again was due to the lack of current to power the LED that was coming from the OR gate. On average the voltage levels of the output of the Z2 OR gate varied from 1.67 Volts to 2.04 Volts.

Table 3 also presents the results when the single-layer was stretched for 37cm to 40cm in the vertical direction when experiments were conducted based on the truth table from Figure 1. Similar challenges to the standard no-stretch tests were encountered. A number of indeterminate readings were occurring, again due to the lack of a high enough threshold voltage coming from the output of the OR gates. As this was not strong enough to power the LED and raise the visual alert notification, it hence left the test in an indeterminate state. When investigated, a number of the voltage output readings for these cases were alternating between 1.6 Volts and 1.9 Volts on average.

Based on the performance results when testing the Multi-layer garment without stretching, the output obtained were as expected according to the defined truth table outputs Z1, Z2. During the tests, a few false outputs were obtained and this was due to a small short circuit that occurred randomly on Z2 OR gate. Upon further investigation, we found that this was due to a small piece of conductive thread that was creating a connection between the base and emitter connection nodes of the T2 transistor. This connection was generating false output readings.

In the case of Multi-layered garment that undergoes stretching of 37cm- 40cm vertically, it was observed that the programmable stitch between the wired bus of the gates in the connector plane had become loose and when stretched it interacted with the circuitry in the vicinity causing the false outputs to occur. It was also noted that the BC connection of the OR gate plane also needed to be tightened to make it more secure. The movement of the garment was having an impact on the stitching interconnecting the two programmable planes. In order to overcome this in a future version, we propose to look at insulating the planes but leaving specific interconnection sites open to allow for the programming of the PLA. This would help secure the circuit and also enhance its overall performance when stretching and moving.

**On-body Edge Computing through E-Textile Programmable Logic Array**

Overall there were more true positives results for the Multi-layer Design as shown in Table 3. In summary, the Single-layer design resulted in high level of indeterminate output results and this is largely due to the lack of stability in this design layout.

## 6    Conclusion

The advancement in textiles and electronic circuitry as well as nanotechnology and material science has resulted in e-textile technology. While the major focus of this technology to date has been on integrating sensing elements seamlessly into the garment, there is still a lack of computing capabilities that is weaved into the garment itself. By integrating the computing functionality into the garment can lead to a new form of On-body Edge Computing. In this paper we have presented the design, development and validation of an e-textile embroidered PLA smart garment. This smart garment enables On-body worn Edge computing capabilities in a smart garment, limiting dependencies on additional devices for their computational ability. The PLA methodology allows the garment to provide a more personalized and safer approach to health monitoring at an individual level. The embroidered PLA garment is extensible to many other use cases across various sectors from healthcare monitoring to monitor health and safety of a workforce. We successfully applied this PLA methodology to an e-textile fabric-based environment leveraging embroidery techniques and conductive threads and validated that it operates, both at a single gate level as well as the full circuitry. In the case of the full PLA circuitry, we initially demonstrated the design and implementation of a one panel fabric implementation of the smart garment PLA for a single-layer design and then a Multi-layered PLA smart garment design. Our results showed that the Multi-layered design achieves more robustness in terms of quantity of voltage that is used, as well as minimized in logic computation errors compared to the Single-layered design, which we found had numerous short circuits. Our proposed solution can usher in a new age of programmable On-body Edge Computing garment for future e-textiles that incorporates both sensing as well as computing on the human body. This in turn can result in new interface designs for 5G/6G wireless interface as well as applications.

## 7    Acknowledgement

D. Henshall and S. Balasubramaniam are funded in part by FutureNeuro from Science Foundation Ireland (SFI) under Grant Number 16/RC/3948 and co-funded under the European Regional Development Fund and by FutureNeuro industry partners. S. Balasubramaniam is also funded by VistaMilk from Science Foundation Ireland (SFI) under Grant Number 16/RC/3835. Frances Cleary is a member LERO, the Irish Software Research Centre supported, in part, by Science Foundation Ireland grant 13/RC/2094. We wish to acknowledge SFI Grant 15/RI/3219 - 'Pervasive Nation'.